\documentclass{cambridge6A}
\usepackage{graphicx}

\usepackage[mathscr]{eucal}
\usepackage{amssymb,latexsym}
\usepackage{verbatim}
\usepackage{amsmath}
\usepackage{amsthm}
\usepackage{enumerate}

\newtheorem{thm}{Theorem}[section]

\newcommand\calE{{\mathcal{E}}}

\newcommand\calT{{\mathcal{T}}}

\newcommand\calY{{\mathcal{Y}}}

\renewcommand\l{\lambda}

\newcommand\bbC{{\mathbb{C}}}

\newcommand\bbT{{\mathbb T}}
\newcommand\bbZ{{\mathbb{Z}}}
\newcommand\bbP{{\mathbb{P}}}

\renewcommand\S{\Sigma}
\newcommand\s{\sigma}
\renewcommand\d{\partial}

\newcommand\f{\phi}
\renewcommand\L{\triangle}
\newcommand\D{\nabla}
\newcommand\e{\epsilon}
\renewcommand\b{\beta}

\newcommand\la{\langle}
\newcommand\ra{\rangle}

\renewcommand\l{\lambda}
\newcommand\g{\gamma}

\renewcommand\a{\alpha}

\renewcommand\th{\theta}

\DeclareFontFamily{OT1}{rsfs}{}
\DeclareFontShape{OT1}{rsfs}{m}{n}{ <-7> rsfs5 <7-10> rsfs7 <10->
rsfs10}{} \DeclareMathAlphabet{\mycal}{OT1}{rsfs}{m}{n}
\def\scri{{\mycal I}}%

\newcommand\<{\la}
\renewcommand\>{\ra}

\newcommand\beq{\begin{equation}}
\newcommand\eeq{\end{equation}}
\newcommand\ben{\begin{enumerate}}
\newcommand\een{\end{enumerate}}
\newcommand\bit{\begin{itemize}}
\newcommand\eit{\end{itemize}}

\newcounter{mnotecount}[section]

\renewcommand{\theequation}{\thesection.\arabic{equation}}
\begin{document}

\chapterauthor{Gregory J. Galloway}

\chapter[Constraints on black hole topology]{Constraints on the topology of higher \\ dimensional black holes}

\contributor{Your name \affiliation{Your affiliation}} 

 

 \section{Introduction}
 \label{intro}
  
As discussed in the first chapter, black holes in four dimensions satisfy remarkable uniqueness properties.  Of fundamental importance is the classical result of Carter, Hawking and Robinson that the Kerr solution, which is characterized by its mass $M$ and angular momentum $J$, is the unique four dimensional asymptotically flat stationary (i.e., steady state)  solution to the vacuum Einstein equations.\footnote{Some recent progress has been made in removing the assumption of analyticity from the classical proof; see e.g., \cite{Alex}.}
A basic step in the proof is Hawking's theorem on the topology of black holes \cite{HE}, which asserts that for such black hole spacetimes, cross sections of the event
horizon are  necessarily spherical, i.e., are topologically $2$-spheres.\footnote{Much later an entirely different proof of this fact was given based on topological censorship, as described in Chapter 1.  However, topological censorship does not in general provide much information about horizon topology in higher dimensions; see the comments in Section \ref{cobordant}.}  In short, for conventional black holes in four dimensions, horizon topology is spherical. 

However, as seen in the previous chapter, in higher dimensions, black hole horizons need not  have spherical topology.     With the remarkable discovery by Emparan and Reall \cite{ER} of the black ring solution, with its  $S^1 \times S^2$ horizon topology, the question naturally arose as to what, if any, are the restrictions on horizon topology in higher dimensional black holes. 
This issue was addressed in a paper of the author and Rick Schoen \cite{GS}, in which we obtained a generalization of Hawking's theorem to higher dimensions.  This generalization is discussed  in Sections \ref{gen} and \ref{proof}.  In preparation for that, we review Hawking's black hole topology theorem in Section \ref{hawking} and introduce some basic background material on marginally trapped surfaces in Section~\ref{mots}.   Theorem~\ref{posyam1} in Section \ref{gen} leaves open the possibility of horizons with, for example, toroidal topology in vacuum black hole spacetimes.  In Section~\ref{borderline} we consider a refinement of Theorem~\ref{posyam1} which rules out such ``borderline"  cases.  In Section \ref{lambda} we address the effect of including the cosmological term in the Einstein equations.
Further constraints on horizon topology are discussed in Section \ref{further}, some based on quite different methods, and some concluding remarks are given in Section~\ref{final}.

 
 \section{Hawking's theorem on black hole topology}
 \label{hawking}

In this section we would like to review Hawking's theorem on black hole topology (as presented in \cite{HE}) and give a brief outline of its proof.  At this point we wish to keep the discussion informal, and hold off on any precise definitions until subsequent sections. 

\begin{thm}[\cite{HE}]\label{HEthm}
Let $M^4$ be a four dimensional asymptotically flat stationary black hole spacetime obeying the dominant energy condition.  Then cross sections of the event horizon are topologically $2$-spheres.
\end{thm}

\begin{figure}[h]
\begin{center}
\mbox{
\includegraphics[width=2.2in]{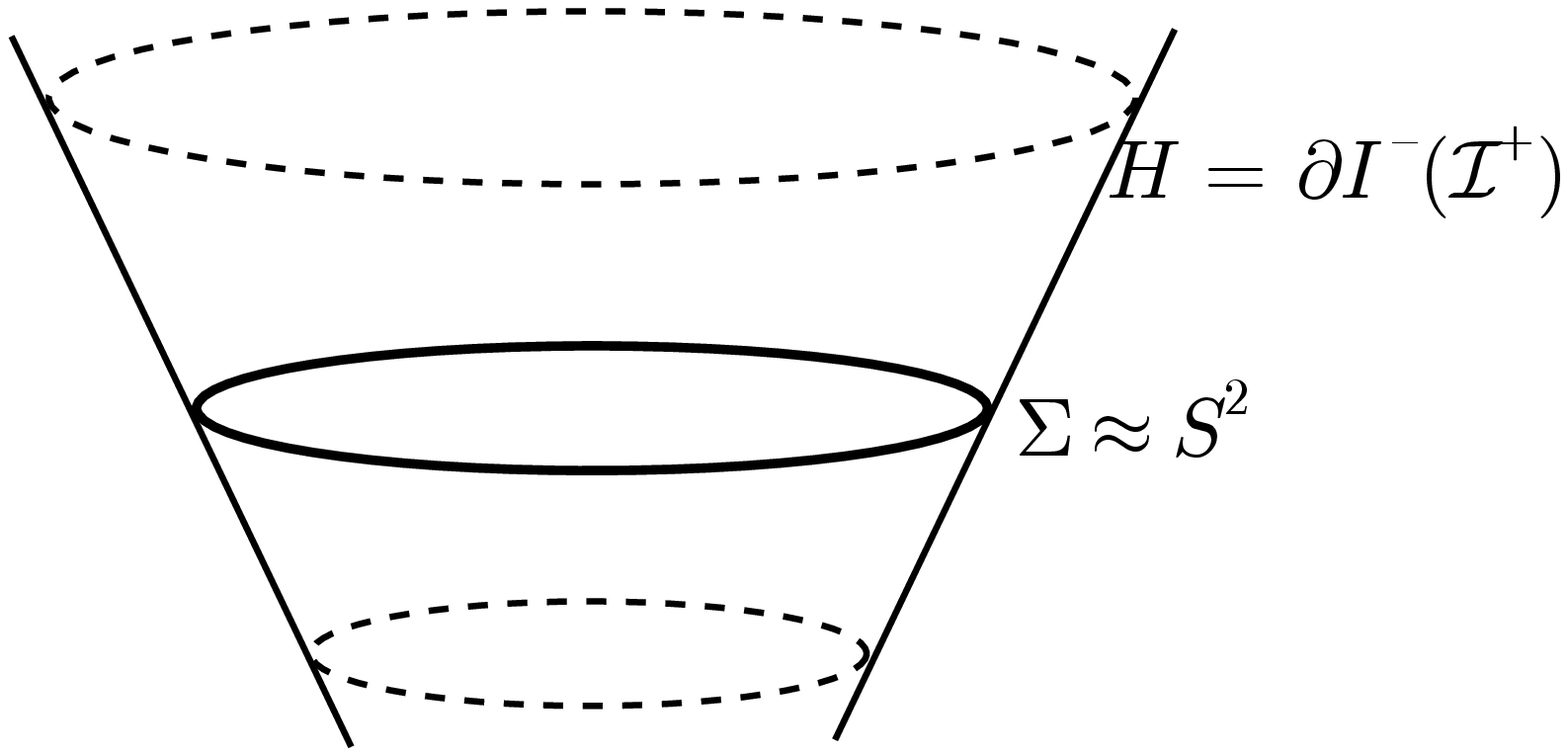}
}
\caption{}
\end{center}
\end{figure}

Here asymptotically flat means that spacetime admits a regular null infinity 
$\scri = \scri^+ \cup \scri^-$.  Then the (future) event horizon is the boundary of the past of future null infinity, $H = \d I^-( \scri^+)$.  By a cross section we mean a smooth compact (without boundary) $2$-surface obtained, say, by intersecting $H$ with a spacelike hypersurface.
As we recall later, the dominant energy condition is a positivity condition on the energy-momentum tensor of the spacetime.  

Apart from one fact, the proof of Theorem \ref{HEthm} is purely local.  Assuming there is a cross section $\S$ that is not spherical, Hawking's proof involves deforming  $\S$ outward to a surface $\S'$ which is {\it outer trapped}, that is to say  the future outward directed
null normal geodesics emanating from $\S'$ are converging along $\S'$.   But it is a basic fact   that outer trapped surfaces cannot occur in the region outside of the black hole; since the 
out-going light rays are converging, 
such surfaces cannot be seen by distant observers, and hence are necessarily contained within the black hole region.  

To construct the outer trapped surface,  Hawking considers a specially chosen one parameter
deformation (or variation) $t \to \S_t$ of $\S = \S_0$ to the past along the null hypersurface generated by the past outward directed normal
null geodesics to $\S$ (see Figure 1.2).   Let $\theta(t)$ denote the expansion of the future outward directed  null normal geodesics emanating from $\S_t$.  
The event horizon $H$ is a null hypersurface ruled by null geodesics, called the null generators  of $H$.    The assumption of stationarity implies that the congruence of  null generators of $H$ has zero expansion.\footnote{By Hawking's area theorem,  this expansion is, in general, nonnegative, but goes to zero in the steady state limit.}  But the future outward directed null normal geodesics of $\S$ coincide with these  generators, and this implies that $\th(0) = 0$.
If $\S$ is not 
a $2$-sphere, and hence has genus (i.e., number of handles) $g \ge 1$, the Gauss-Bonnet theorem
and dominant energy condition are then used to show that $\left . \frac{\d\th}{\d t} \right |_{t=0} < 0$. It follows that for sufficiently small $t >0$, $\th(t) < 0$, which implies that $\S_t$ is outer trapped.  Hence, $\S$ must be a $2$-sphere.

\begin{figure}[h]
\begin{center}
{\hspace*{.3in}\mbox{
\includegraphics[width=2.2in]{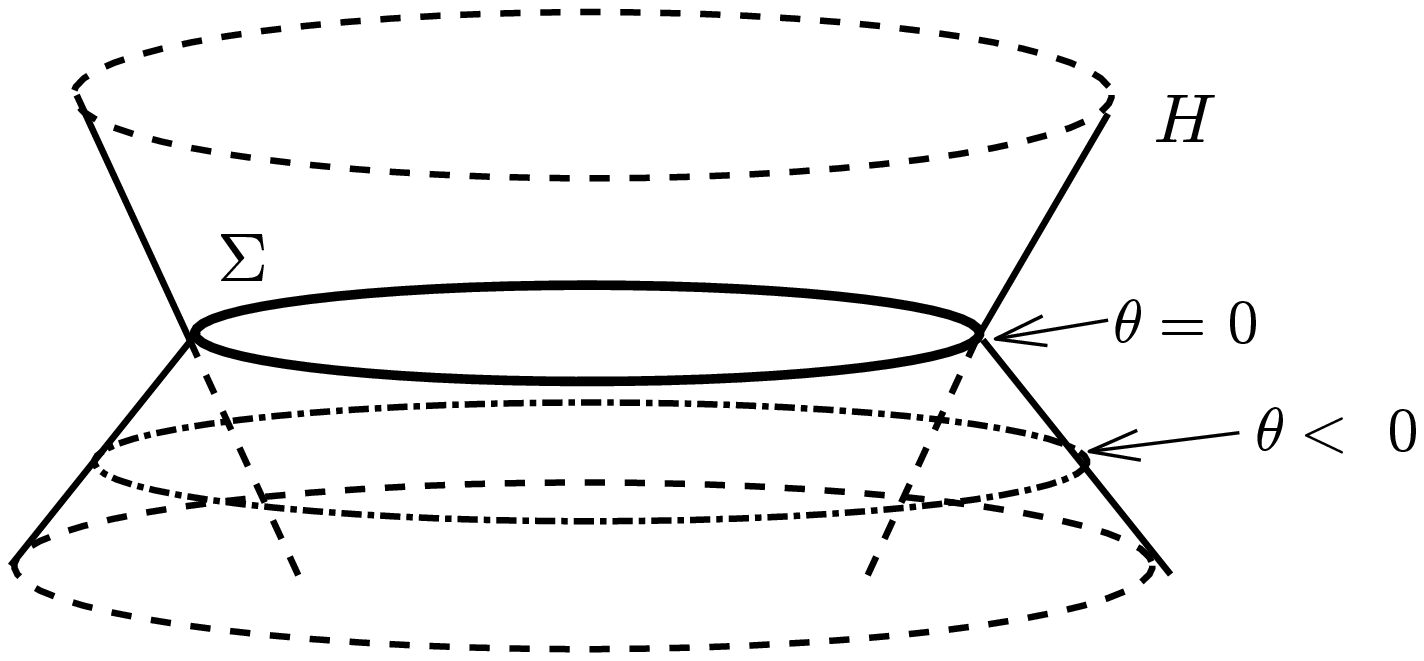}
}}
\caption{}
\end{center}
\end{figure}

Actually, the torus $\bbT^2$ ($g=1$) arises as a borderline case in the proof.  The arguments  in \cite{HE} show that the torus could arise only under special 
circumstances, e.g., $\S$ would have to be flat and a certain energy-momentum tensor term would have to vanish along $\S$.   It is not quite clear to us, though, that the arguments succeed in eliminating the possibility of a torus altogether.
In any case, as the proof relies on the Gauss-Bonnet theorem, it does not directly generalize  to higher dimensions.  

In \cite{Hawking}, Hawking showed how to extend his black hole topology result to {\it apparent horizons}, i.e., to outermost marginally outer trapped surfaces.  Here, `outermost' is with respect to a given spacelike hypersurface.   Our generalization of Hawking's theorem is carried out in this more general context.  Moreover, like Hawking's proof, our proof is variational in nature.
 
\section{Marginally outer trapped surfaces} 
\label{mots}

The notion of a marginally outer trapped surface  was introduced early on in the development of the theory of black holes.  Under suitable circumstances, the occurrence of a marginally outer trapped surface in a time slice of spacetime signals the presence of a black hole \cite{HE}.   For this and other reasons marginally outer trapped surfaces have played  a fundamental role in quasi-local descriptions of  black holes; see e.g.,  \cite{AK}.   They also play an important role in numerical simulations of black hole formation, black hole collisions, etc., and many numerical algorithms have been developed to find them. The mathematical theory of  marginally outer trapped surfaces has been broadly developed in recent years, see e.g. the 
recent survey article \cite{AEM}.  

Let $(M^{n+1},g)$ be a spacetime (time oriented Lorentzian manifold) of dimension $n+1$, $n \ge 3$.   Let $V^n$ be a spacelike hypersurface in $M^{n+1}$, with induced metric $h$ and second fundamental form (extrinsic curvature tensor) $K$.  Thus, for tangent vectors $X, Y$ to $V$ at a given point, $K(X,Y) = g(\D_X u,Y) = X^{\mu}Y^{\nu}\D_{\mu}u_{\nu}$, where $\D$ is the Levi-Civita connection of $M$ and $u$ is the future directed timelike unit normal vector field to $M$. 

Let $\S^{n-1}$ be a compact  hypersurface in $V^n$, and assume that 
$\S$ separates $V$ into an ``inside" and ``outside"; let $\nu$ be the outward pointing unit normal vector field to $\S$ in $V$.  Then $l_+ = u+\nu$ (resp. $l_- =  u - \nu$) is a future directed outward (resp., future directed inward) pointing null normal vector field along $\S$, unique up to positive scaling (see Figure 1.3).
\begin{figure}[b]
\vspace*{.1in}
\begin{center}
\mbox{
\includegraphics[width=2.9in]{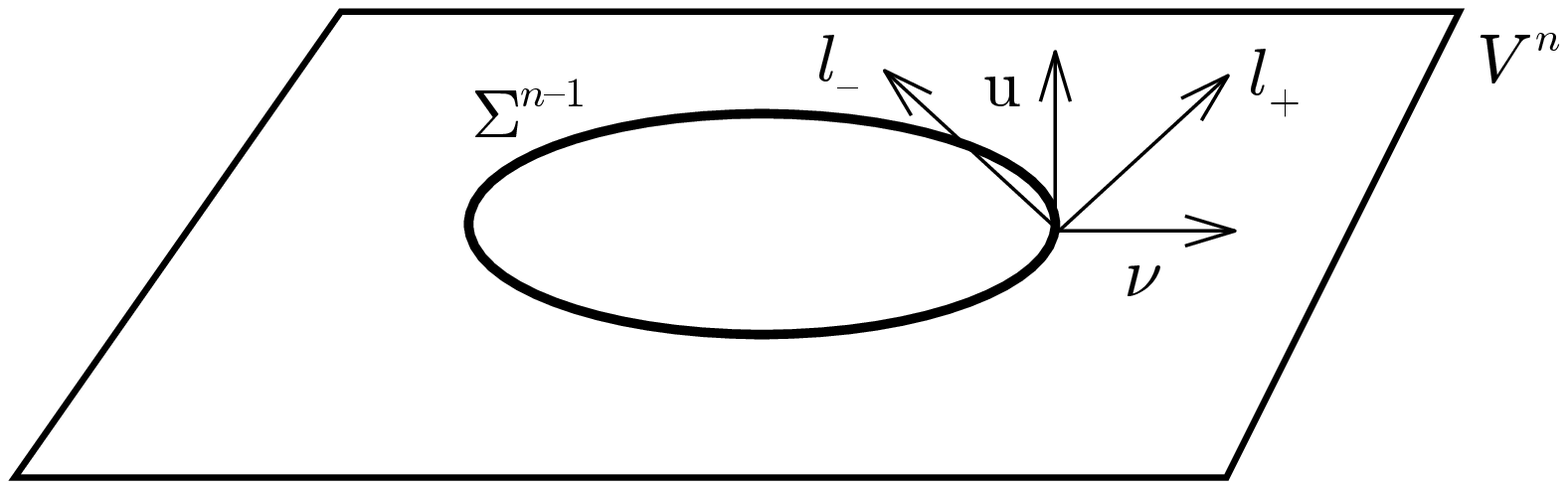}
}
\caption{}
\end{center}
\end{figure}

The second fundamental form of $\S$, viewed as a submanifold of spacetime, can be decomposed into two scalar valued {\it null second forms},  $\chi_+$ and $\chi_-$, associated 
to $l_+$ and $l_-$, respectively.
At each point of $\S$, $\chi_{\pm}$ is 
the bilinear form defined by,
\beq
\chi_{\pm}(X,Y) = g(\D_Xl_{\pm}, Y) = X^{\mu}Y^{\nu}\D_{\mu}(l_{\pm})_{\nu} \,.
\eeq
for pairs of tangent vectors $X,Y$ to $\S$.
The {\it null expansion scalars}   $\th_{\pm}$ of $\S$   are obtained by tracing 
$\chi_{\pm}$ with respect to the induced metric $\g$ on $\S$,
\begin{align}
\theta_{\pm} = {\rm tr}_{\g} \chi_{\pm} = \g^{AB}(\chi_{\pm})_{AB} = {\rm div}\,_{\S} l_{\pm}  \,.
\end{align}
It is easy to check that the sign of $\th_{\pm}$ is invariant under positive rescaling 
of the null vector field $l_{\pm}$. The vector fields $l_{\pm}$ correspond to the initial
tangents of the future directed null geodesics issuing orthogonally from $\S$. Thus, physically, 
$\th_+$ (resp., $\th_-$) measures the divergence of the  outgoing (resp., ingoing) light rays emanating from $\S$.  One can express the null expansion scalars in  terms of the {\it initial data} $h, K$ on $V^n$, as follows,
\beq\label{thid}
\th_{\pm} = {\rm tr}_{\g} K \pm H \,,
\eeq 
where $H$ is the mean curvature of $\S$ within $V$.  Note in particular, in the time-symmetric case, $K \equiv 0$, $\th_+$ is just the mean curvature of $\S$.

For round spheres in Euclidean slices of Minkowski space, with
the obvious choice of inside and outside, one has $\th_- < 0$
and $\th_+ >0$. In fact, this is the case in general for large
``radial" spheres in {\it asymptotically flat} spacelike
hypersurfaces.   However, in regions of space-time where the
gravitational field is strong, one may have both $\th_- < 0$
and $\th_+ < 0$, in which case $\S$ is called a {\it trapped
surface}.    Under appropriate energy and causality conditions, 
the occurrence of a trapped surface signals  the onset of gravitational collapse (this is the implication of the Penrose singularity theorem) and the existence
of a black hole \cite{HE}.

Focusing attention on the outward null normal only, we say that
$\S$ is an outer trapped surface  if $\th_+ < 0$.  Finally, we define $\S$ to be a marginally
outer trapped surface (MOTS) if $\th_+$ vanishes identically.

MOTSs arise naturally in a number of situations.  Most basically, as pointed out in our discussion of the proof of Hawking's black hole topology theorem, cross sections of the event horizon (obtained, say, as the smooth compact intersection of the event horizon with a spacelike hypersurface),  in stationary  black holes spacetimes  are MOTSs.  

\begin{figure}[t]
\vspace*{.1in}
 \centering
  \begin{minipage}[b]{5.5 cm}
\includegraphics[width=2.4in]{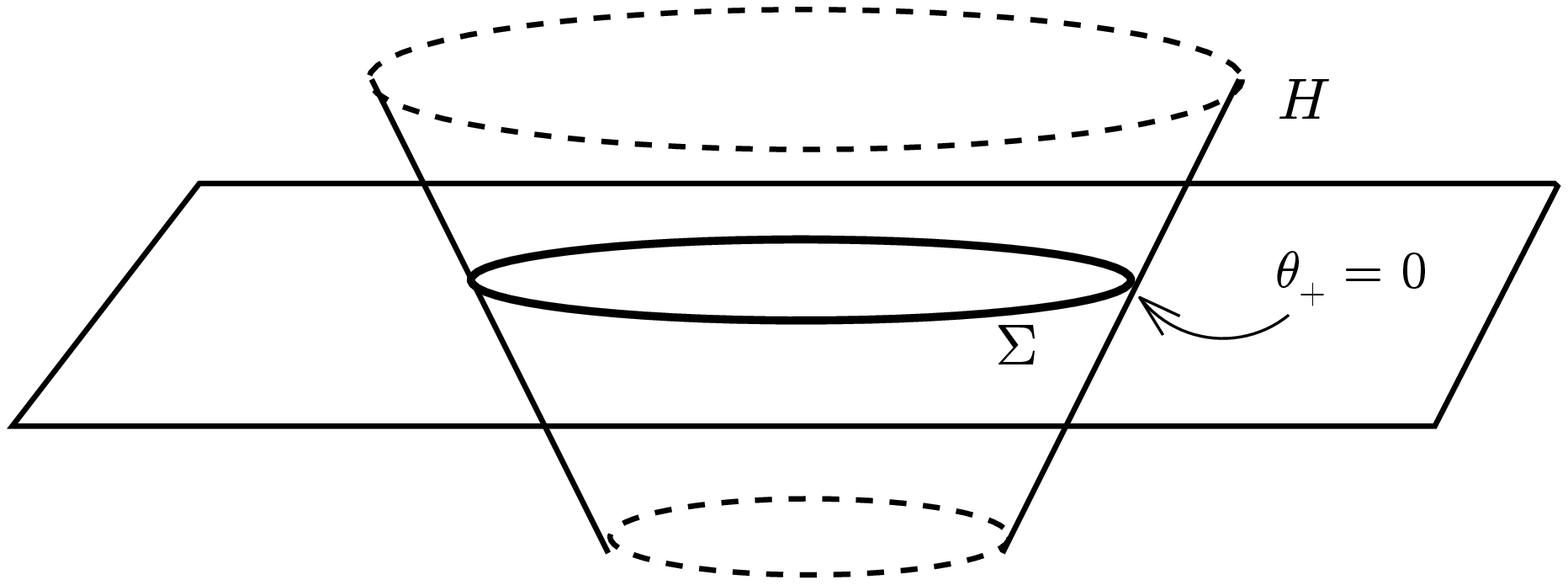}
  \end{minipage}
  \begin{minipage}[b]{5.5 cm}
 \includegraphics[width=2.4in]{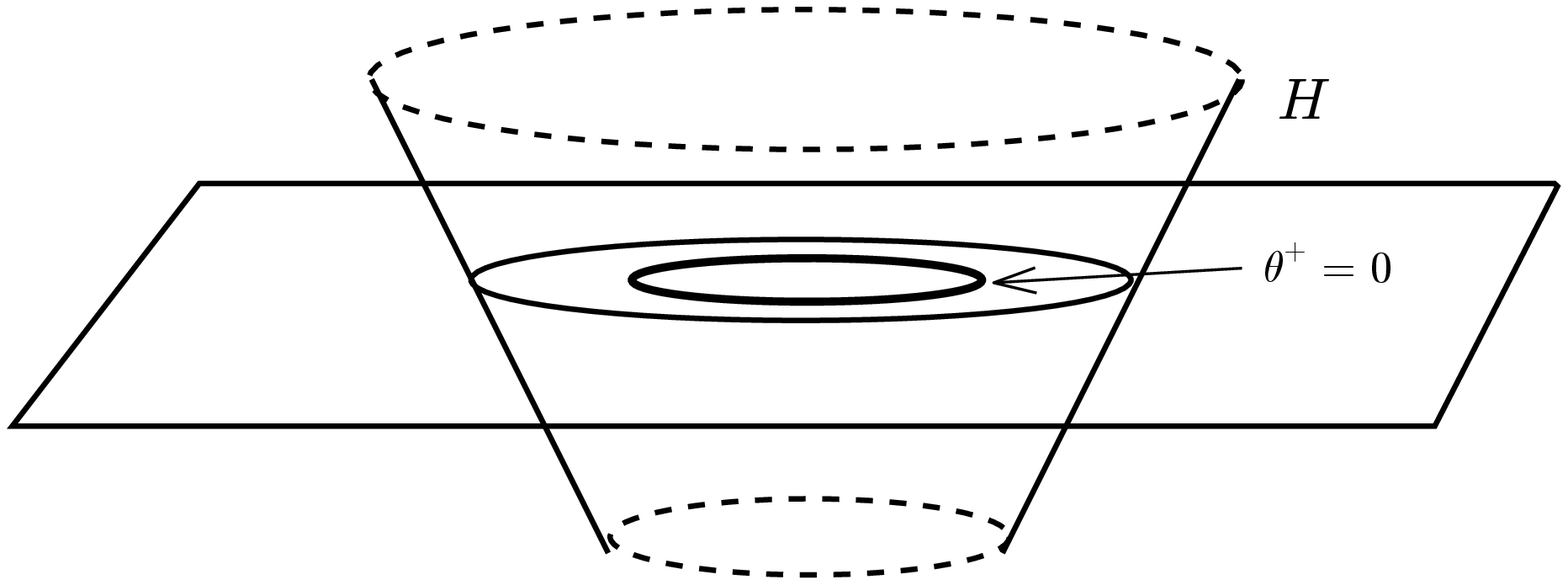}
  \end{minipage}
  \caption{In stationary black hole spacetimes (figure at left), 
  cross sections of the event horizon are MOTSs.   In dynamical black hole spacetimes 
(figure at right), MOTSs typically occur inside the black hole.}
\end{figure}

In dynamical black hole spacetimes MOTSs typically occur inside the event horizon. (In fact they are forbidden to occur outside the black hole.)  There are old heuristic arguments for the existence of MOTSs in this case, based on considering the boundary of the trapped region inside the event horizon.  These heuristic ideas have recently been
made rigorous first  by Andersson and Metzger \cite{AM2}  for three dimensional initial data sets, and then by Eichmair \cite{Eich1,Eich2}  for initial data sets up to dimension seven.  These results rely on a basic existence result for MOTSs under physically natural barrier conditions, and imply the existence of outermost MOTSs as described in the next section. We refer the reader to the  survey article \cite{AEM} for an excellent discussion of how the existence of MOTSs is established by inducing blow-up of Jang's equation.   

\section[A generalization of Hawking's Theorem and some topological restrictions]{A generalization of Hawking's Theorem and some \\ topological restrictions}
\label{gen}

Let $V^n$ be a spacelike hypersurface in the spacetime $(M^{n+1}, g)$, $n \ge 3$, as in the previous subsection.   Henceforth, assume that spacetime satisfies the Einstein equations (without
cosmological constant),
$$
{\rm Ric} - \frac12 R g = \calT  \,.
$$
Then, $M$ is said to obey the {\it dominant energy condition} (DEC) provided the energy-momentum tensor
$\calT$ satisfies, $\calT(X,Y) = T_{\mu\nu} X^{\mu}Y^{\nu} \ge 0$ for all future directed causal vectors $X,Y$.

Our generalization of Hawking's theorem applies to outermost MOTSs.   We say that a MOTS 
$\S$ in $V$ is outermost provided there are no outer trapped ($\th_+ < 0$) or marginally outer trapped ($\th_+ = 0$) surfaces outside of, and homologous to, $\S$.\footnote{Here, ``\,$\S'$  homologous to $\S$\," simply means  that $\S$ and $\S'$ form the boundary of a compact region in $V$. We don't care about the occurrence of outer trapped surfaces or MOTSs outside of $\S$ that are not homologous to $\S$.}   It is a fact that a cross-section  $\S$ of the event horizon in an asymptotically flat black hole spacetime obeying the 
DEC\,\footnote{Actually the {\it null energy condition}, ${\rm Ric}(X,X) = R_{\mu\nu}X^{\mu}X^{\nu} \ge 0$ for all null vectors $X$,  suffices for this.}
is an outermost MOTS relative to
any  spacelike hypersurface whose intersection with the horizon is $\S$.  Again, this is
because outer trapped surfaces, or even marginally outer trapped surfaces homologous to $\S$ cannot occur outside the black hole region (see Figure 1.5).   

\begin{figure}[t]
\vspace*{.1in}
\begin{center}
\mbox{
\includegraphics[width=3.3in]{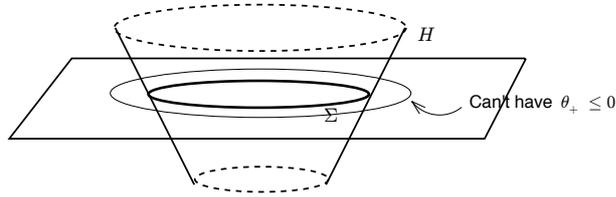}
}
\caption{Cross sections of the event horizon in asymptotically flat stationary black hole spacetimes obeying the DEC are outermost MOTSs.}
\end{center}
\end{figure}

More generally, results of Andersson and Metzger \cite{AM2} (in three spatial dimensions) and
Eichmair \cite{Eich1,Eich2} (up to seven spatial dimensions) guarantee the existence of outermost 
MOTSs under natural barrier conditions.  
More specifically, suppose $\S_1$ is an outer trapped
surface in $V^n$, $3 \le n \le 7$, and suppose there is a surface $\S_2$ outside of and homologous to $\S_1$ which is outer {\it untrapped}, i.e., which has outer null expansion $\th_+ > 0$.  (For example $\S_2$ might be a large sphere out near infinity on
an asymptotically flat end of $V$.)  Then the results of Andersson-Metzger and  of Eichmair imply the existence  of an outermost MOTS  $\S$ in the region bounded by $\S_1$ and $\S_2$ (see Figure 
1.6).  For further details, see \cite{AEM} and references therein.  
\begin{figure}[h]
\begin{center}
\mbox{
\includegraphics[width=2.4in]{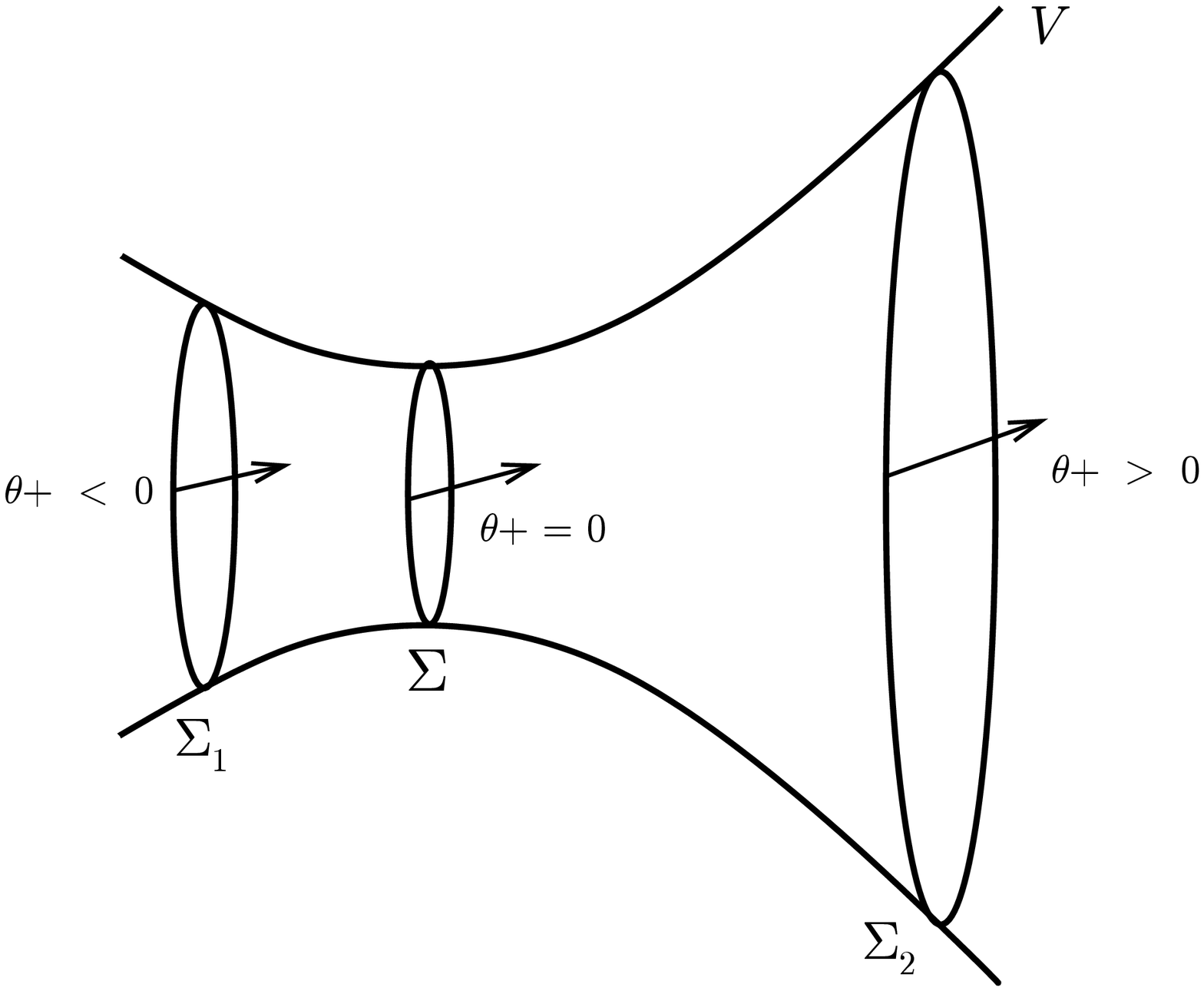}
}
\caption{}
\end{center}
\end{figure}

We need to introduce one last piece of terminology.   A smooth compact manifold 
is said to be of {\it positive Yamabe type} if it admits a Riemannian metric of 
positive scalar curvature.  By the solution of the Yamabe problem,
the conformal class of this metric will contain 
a metric of {\it constant} positive scalar curvature,
but we shall not need that fact here.   


We are now ready to state the generalization of Hawking's theorem.

\begin{thm}[\cite{GS}]\label{posyam1}
Let $V^n$, $n \ge 3$, be a spacelike hypersurface in a spacetime obeying the DEC. 
If $\S^{n-1}$ is an outermost MOTS  in $V^n$ 
then 
$\S^{n-1}$ is of positive Yamabe type, \underline{unless}
\,$\S$ is Ricci flat (flat if $n =3,4$),
$\chi_+ \equiv 0$, and 
$\calT(u,l_{+}) =$ $T_{\mu\nu}u^{\mu} l_+^{\nu} \equiv 0$ on $\S$.
\end{thm}

Thus, apart from certain exceptional circumstances (which we ignore for now, but will address later), $\S$ is of positive Yamabe type.  The relevance of this for black hole topology is that there
is an extensive literature concerning results which establish restrictions on the topology of manifolds that admit metrics of positive scalar curvature.   We consider two basic examples now, and discuss some further restrictions in Section \ref{obstruct}.  
For simplicity, in the present discussion we assume $\S$ is orientable.

\smallskip
\noindent
\underline{Case 1}.  $\dim \S =2$ ($\dim M = 3+1$).  In this case, $\S$ 
being of positive Yamabe type  means that $\S$ admits a metric of positive Gaussian curvature.  Hence, by the Gauss-Bonnet theorem, $\S$ is topologically  a $2$-sphere, and we recover Hawking's theorem.

\smallskip
\noindent
\underline{Case 2}.  $\dim \S =3$ ($\dim M = 4+1$).   In this case we have the following
result. 

\begin{thm}\label{3dim}
If $\S$ is a compact orientable
$3$-manifold of positive Yamabe type then $\S$ must be diffeomorphic to (i)
a spherical space, or (ii) $S^1 \times S^2$, or
(iii) a connected sum of  the previous two types.
\end{thm}
  
By a spherical space we mean the $3$-sphere $S^3$ or, more generally, a space covered by $S^3$, such as a lens space.  Thus, the basic horizon topologies in the case
$\dim \S =3$ are $S^3$ and $S^1 \times S^2$, with the latter being realized by the
black ring.  

The proof of Theorem \ref{3dim}  goes briefly as follows.  By the prime decomposition 
theorem~\cite{Hempel}, $\S$ can be expressed as a connected sum of (i) spaces covered
by homotopy $3$-spheres, (ii) $S^1 \times S^2$'s and (iii) $K(\pi,1)$ spaces.  We recall that a $K(\pi, 1)$ space is a space whose universal cover is contractible, such as the $3$-torus.  Now, by a result of Gromov and Lawson \cite{GL3}, a manifold that admits a metric of positive scalar curvature cannot
have any $K(\pi,1)$'s in its prime decomposition.  Moreover, by the positive resolution of the
Poincar\'e conjecture, the only homotopy $3$-sphere is the  $3$-sphere.  The theorem follows.

All the $3$-manifolds listed in Theorem \ref{3dim} admit metrics of positive scalar curvature, but 
so far only the $S^3$ and $S^1 \times S^2$ topologies have been realized by asymptotically flat
stationary black hole spacetimes obeying the Einstein equations.   Further restrictions on the horizon topology have been obtained under the assumption of additional symmetries.  For example, in \cite{Hollands1} it is shown that  for asymptotically flat, stationary vacuum black holes in five dimensions with two commuting axial symmetries,  the horizon must be topologically either a $3$-sphere, an $S^1 \times S^2$, or a lens space.  If there is only one axial symmetry, which is guaranteed to exist for analytic asymptotically flat stationary vacuum black holes \cite{Hollands0, Isenberg}, some restrictions on the horizon topology can still be obtained beyond Theorem \ref{3dim}.  Roughly, in this case it is shown in \cite{Hollands2} that the horizon either is a connected sum of  lens spaces and $S^1 \times S^2$'s (with at least one $S^1 \times S^2$ present) or one of several possible quotients of $S^3$ by isometries.   
Topological censorship \cite{FSW, GSSW, CGS} and certain techniques
used in our proof of Theorem~\ref{posyam1}  are two of the ingredients used in their proof, which
involves a detailed analysis of the quotient of a horizon cross-section by the $U(1)$ action.


\section{The proof of Theorem \ref{posyam1}}
\label{proof}

Let the setting be as in the statement of Theorem \ref{posyam1}.  As noted earlier, the proof, like Hawking's, is variational in nature.   
We consider a one-parameter
deformation (or variation) $t \to \S_t$  of $\S = \S_0$ with initial deformation velocity 
$v= \left . \frac{\d}{\d t}\right |_{t=0} = \phi \nu$, where, recall, $\nu$ is the outward pointing unit
normal to $\S$ in $V$, and $\phi$ is a smooth function on $\S$.  Such a deformation can be 
achieved by, for each $x \in \S$, moving along the geodesic starting at $x$ with initial velocity 
$\phi \nu|_x$ a time $t$.  For $t$ sufficiently small, this produces a smooth variation of~$\S$ (see Figure 1.7).
\begin{figure}[b]
\begin{center}
\mbox{
\includegraphics[width=3.5in]{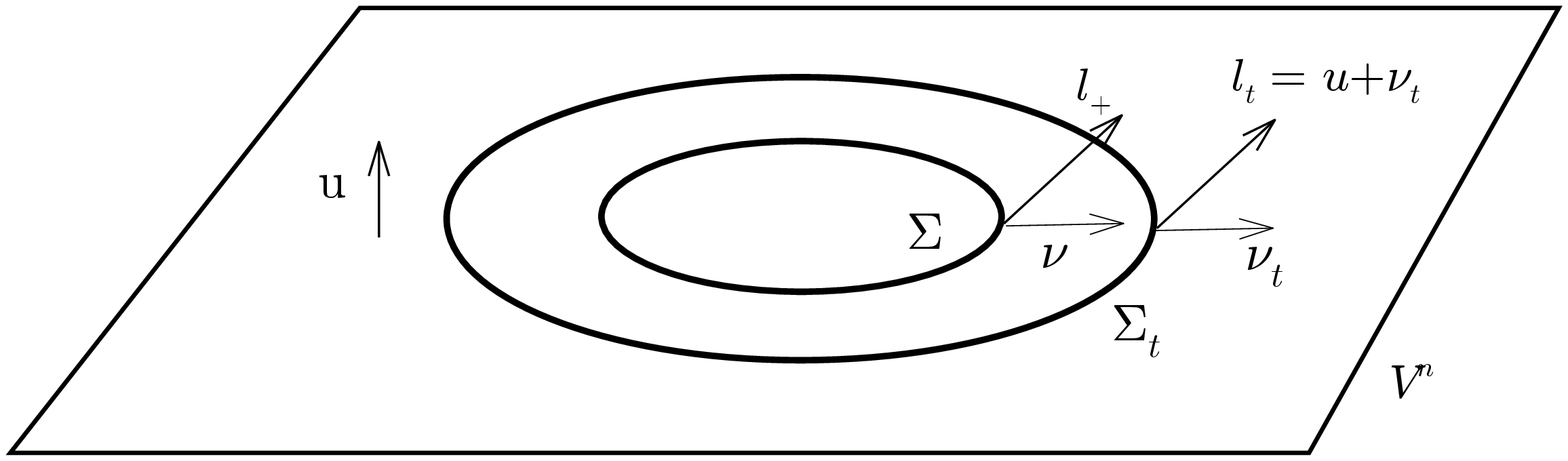}
}
\caption{}
\end{center}
\end{figure}

Let $\th(t)$ denote
the null expansion of $\S_t$ with respect to $l_t = u + \nu_t$, where $\nu_t$ is the
outward  unit normal field to $\S_t$ in $V$.  A computation shows \cite{AMS2},
\beq\label{der}
\left . \frac{\d\th}{\d t} \right |_{t=0}   = L(\f) \;, 
\eeq 
where $L : C^{\infty}(\S) \to C^{\infty}(\S)$ is
the operator, 
\beq\label{stabop}
L(\phi)  = -\triangle \phi + 2\<X,\D\phi\>  + \left( \frac12 S
- \calT(u,l_+) - \frac12 |\chi_+|^2+{\rm div}\, X - |X|^2\right)\phi \,. 
\eeq 
In the above, $\triangle$, $\D$ and ${\rm
div}$ are the Laplacian, gradient and divergence operator,
respectively, on $\S$, $S$ is the scalar curvature of $\S$, $X$ is the 
vector field  on $\S$  dual to the one form $K(\nu,\cdot)|_{\S}$, and
$\<\, \cdot ,\, \cdot \>$ denotes the induced $\g$ metric  on $\S$.  
$L$ is referred to as the stability operator associated with variations in the null expansion $\th$.
In the time symmetric case considered in \cite{CG}, the vector field $X$ vanishes, and $L$ reduces to the classical stability operator of minimal surface theory.

Now consider the eigenvalue problem,
\beq
L(\f) = \l \f  \,. 
\eeq
$L$ is a second order linear elliptic operator,  which, owing to the first order term, is
not in general self-adjoint.  As such it may have  some nonreal eigenvalues.  Nevertheless, as discussed in \cite{AMS1, AMS2}, its {\it principal eigenvalue} $\lambda_1$ (i.e., the eigenvalue with smallest real part) is necessarily real, and, moreover, one can choose a principal eigenfunction
$\phi$ (hence satisfying, $L(\f) = \l_1 \f$) which is strictly positive, $\phi > 0$.  Using the eigenfunction $\phi$ to
define our variation, we have from (\ref{der}),
\beq\label{der3}
\left . \frac{\d\th}{\d t} \right |_{t=0} =\l_1 \phi \,.
\eeq
The eigenvalue $\l_1$ cannot be negative, for otherwise (\ref{der3}) would
imply that  $\frac{\d\th}{\d t}< 0$ on $\S$.  Since $\th = 0$ on $\S$, this would mean
that for $t>0$ sufficiently small, $\S_t$ would be outer trapped, contrary to our
assumption that $\S$ is an outermost MOTS.

Hence, $\l_1 \ge 0$, and we conclude for the variation determined by the
positive eigenfunction $\phi$ that $\left . \frac{\d\th}{\d t} \right |_{t=0} = L(\f) \ge 0$.
By completing the square on the right hand side of Equation (\ref{stabop}), this implies
that the following  inequality holds,
\beq
 -\triangle \phi +\left(Q+{\rm div}\, X  \right)\phi   +
\phi|\D \ln\phi|^2  - \phi|X - \D\ln\phi|^2  \ge 0 \,,
\eeq
where, for notational convenience, we have put,
\beq\label{Q}
Q =   \frac12 S - \calT(u,l_+) - \frac12 |\chi_+|^2  \,.
\eeq
Setting $u = \ln \phi$, we obtain,
\beq\label{u-ineq}
-\triangle u +Q + {\rm div}\,X  - |X - \D u|^2 \ge 0 \,.
\eeq

As a side remark,  note that integrating  this inequality gives that the total scalar curvature
of $\S$ is nonnegative, and in fact is positive, except under special circumstances. 
In four spacetime dimensions one may then apply the Gauss-Bonnet theorem to recover Hawking's theorem; in fact this is essentially Hawking's original argument.  However, in higher
dimensions the positivity of the total scalar curvature, in and of itself,  does not provide any topological information.  

To proceed, we first absorb the Laplacian term
$\triangle u = {\rm div}\,(\D u)$ in (\ref{u-ineq})  into the divergence term to
obtain,
 \beq
Q + {\rm div}\,(X- \D u)  - |X - \D u|^2 \ge 0  \, .
\eeq
Setting $Y = X - \D u$, we arrive at the inequality,
\beq
- Q + |Y|^2 \le {\rm div}\, Y  \,.
\eeq

Now comes a simple, but critical estimate, of a sort first considered in \cite{SY2}.
Given any $\psi \in C^{\infty}(\S)$, we multiply  through by $\psi^2$ and derive,
\begin{align}
-\psi^2 Q +\psi^2 |Y|^2 & \le \psi^2 {\rm div}\, Y \nonumber\\
& =  {\rm div}\,(\psi^2Y) - 2\psi \< \D\psi,Y \>  \nonumber  \\
& \le   {\rm div}\,(\psi^2Y) + 2|\psi| |\D \psi| |Y| \nonumber\\
& \le  {\rm div}\,(\psi^2Y) + |\D \psi|^2 + \psi^2|Y|^2  \, .
\end{align}
Integrating the above inequality yields,
\beq\label{psi-ineq}
\int_{\S} |\D \psi|^2 + Q \psi^2 \ge 0 \quad \mbox{for all } \psi \in C^{\infty}(\S)  \,,
\eeq 
where $Q$ is given in (\ref{Q}).

Now consider the eigenvalue problem,
\beq\label{eigen}
\hat L(\f) = \l \f
\eeq
where $\hat L : C^{\infty}(\S) \to C^{\infty}(\S)$ is the second order linear elliptic  operator,
\beq\label{op}
\hat L(\f) =  - \triangle\f + Q\, \f \,,
\eeq
obtained formally from \eqref{stabop} by setting $X = 0$.  For self-adjoint operators of the form
\eqref{op}, the Rayleigh formula \cite{Evans} and an integration by parts gives the following standard characterization of the principle eigenvalue $\hat \l_1$ of $\hat L$,
\begin{align}\label{ray}
\hat \l_1 
&= \inf_{\psi \not\equiv 0} \frac{\int_{\S} \psi \hat L(\psi) \,d\mu}{\int_{\S} \psi^2 \,d\mu} \nonumber  \\ 
&= \inf_{\psi \not\equiv 0} \frac{\int_{\S} |\D\psi|^2 + Q \psi^2 \,d\mu}{\int_{\S} \psi^2 \,d\mu}  \,.
\end{align}

It now follows from \eqref{psi-ineq} that $\hat \l_1 \ge 0$.  At this stage, the proof has reduced
{\it morally} to the time-symmetric case as considered in \cite{CG}, and the remainder of the argument can proceed in a similar fashion.  

Let $\hat \f$ be an eigenfunction associated to $\hat \l_1$; $\hat \f$ can be chosen to be strictly
positive, $\hat \f > 0$. Consider $\S$ in the conformally related metric 
$\hat \g = \hat \f^{2/n-2} \g$.  By a standard formula for conformally related metrics,
the scalar curvature $\hat S$ of $\S$ in the metric
$\hat \g$ is given by,
\begin{align}\label{scalar}
\hat S & =  \hat \f^{-n/(n-2)}\left (-2\L\hat \f + S \hat \f + \frac{n-1}{n-2} \frac{|\D\hat \f|^2}{\hat \f}\right)
\nonumber  \\
& =  \hat \f^{-2/(n-2)}\left (2\hat \l_1  + 2\calT(u,l_+) + |\chi_+|^2 + \frac{n-1}{n-2} 
\frac{|\D \hat \f|^2}{\hat \f^2} \right)\,,
\end{align}
where, for the second equation, we have used (\ref{eigen})-(\ref{op}), with $\f =\hat  \f$, and 
(\ref{Q}).

Since, by the dominant energy condition, $\calT(u,l_+) \ge 0$, Equation (\ref{scalar})
implies that $\hat S \ge 0$.  If $\hat S > 0$ at some point, then by well known results
\cite{KW} one can conformally change $\hat \g$  to a metric 
of strictly positive scalar curvature, and the theorem follows.  If $\hat S$ vanishes
identically then, by Equation (\ref{scalar}), $\hat \l_1 = 0$, $\calT(u,l_+) \equiv 0$, 
$\chi_+ \equiv 0$ and $\hat \f$ is constant.  Equations (\ref{eigen})-(\ref{op}), with 
$\f = \hat \f$ and Equation
(\ref{Q}) then  imply that $S \equiv  0$.  By a result of Bourguinon (see \cite{KW}),
it follows that $\S$ carries a metric of positive scalar curvature unless  it is
Ricci flat.  Theorem \ref{posyam1} now follows.

\smallskip
\noindent
\underline{\it Remark:}  With regard to the assumption that $\S$ is an outermost MOTS, the
proof shows that it is sufficient to assume the existence of  a {\it positive} ($v = \f \nu$, with $\f >0$) variation
$t \to \S_t$ such that $\left . \frac{\d\th}{\d t} \right |_{t=0} \ge 0$.  Such a MOTS is called stable
in \cite{AMS1, AMS2}, and arguments in \cite{AMS1,AMS2} show that $\S$ is stable if and only if $\l_1 \ge 0$,
where $\l_1$ is the principal eigenvalue of the operator $L$ given in \eqref{stabop}.  So, in other words, in Theorem \ref{posyam1}, it is sufficient to assume the MOTS $\S$ is stable.

\section{The borderline case}
\label{borderline}

A drawback of Theorem \ref{posyam1} is that, when the DEC along $\S$ does not hold strictly, it allows certain possibilities that one would like to rule out.  For example, it does not rule out the possibility of a vacuum black hole spacetime with toroidal horizon  topology.  Eventually, we were able
to remove the exceptional case (the ``unless" clause) in Theorem \ref{posyam1} altogether, and hence prove the following.  

\begin{thm}[\cite{G}]\label{posyam2}
Let $V^n$, $n \ge 3$, be a spacelike hypersurface in a spacetime obeying the DEC.   If $\S^{n-1}$ is an outermost MOTS  in $V^n$ 
then $\S^{n-1}$ is of positive Yamabe type.
\end{thm}

Thus, without exception, cross sections of the event horizon in asymptotically flat stationary  black hole spacetimes obeying the dominant energy condition are of positive Yamabe type.  In particular,
there can be no toroidal horizons.   

We remark that it is not sufficient to assume in Theorem \ref{posyam2} that $\S$ is stable, in the sense described at the end of Section \ref{proof}; there are counter examples in this case.

Theorem \ref{posyam2} is an immediate consequence of  the following rigidity result.  

\begin{thm}[\cite{G}]\label{stab}
Let $V^n$, $n \ge 3$, be a spacelike hypersurface in a spacetime obeying the DEC.
Suppose  $\S$ is a MOTS in $V$ such that there are
no outer trapped surfaces ($\th_+ < 0$) outside of, and homologous, to $\S$.  If $\S$ 
is NOT of positive Yamabe type,  then there exists
an outer neighborhood $U \approx [0,\e) \times \S$ of $\S$ in $V$ such that each slice
$\S_t = \{t\} \times \S$, $t \in [0,\e)$ is a MOTS.
\end{thm}
 
Thus, if $\S$ is not of positive Yamabe type, there would have to exist either an outer trapped or marginally outer trapped surface outside of and homologous to $\S$, and hence $\S$ would not be outermost.  

We make a brief comment about the proof of Theorem \ref{stab}.  The proof
consists of two steps.  In the first step, one uses Theorem \ref{posyam1} and an inverse function theorem argument to obtain an outer foliation $t \to \S_t$, $0 \le t \le \e$, of surfaces $\S_t$ of {\it constant} outer null expansion, $\th(t) = c_t$.  The second step involves showing that the constants $c_t = 0$.  This latter step requires a reduction to the case that $V$ has nonpositive mean curvature  near 
$\S$, which is achieved by a small spacetime deformation of $V$ in a neighborhood of
$\S$.   The proof makes use of the formula for the $t$-derivative, $\frac{\d\th}{\d t}$, not just at $t=0$ where $\th =0$, but all along the foliation $t \to \S_t$, where, a priori, $\th(t)$ need not be zero.  Thus,  additional terms appear in the expression for $\frac{\d\th}{\d t}$ beyond  those in \eqref{der}-\eqref{stabop}, including a term involving the mean curvature of $V$, which need to be accounted for.  See \cite{G} for details.

\section{Effect of the cosmological constant}
\label{lambda}

 Now suppose that spacetime obeys the Einstein equations with cosmological term,
 \beq\label{eeq+}
{\rm Ric} - \frac12 R g = \calT - \Lambda g  \,.
\eeq
What is sufficient for the proof of Theorem \ref{posyam1} is that the effective energy-momentum tensor $\calT' = \calT - \Lambda g$ satisfy,
$$
\calT'(u, l_+) = \calT(u,l_+) + \Lambda \ge 0 \,. 
$$
In particular, if the fields contributing to $\calT$ obey the DEC and $\Lambda \ge 0$ then Theorem
\ref{posyam1} remains valid (similarly for Theorem \ref{posyam2}). 

However, if $\Lambda < 0$ then the {\it effective} DEC may fail to hold, and in this case Hawking's arguments and the generalization of those arguments presented here do not yield any topological
conclusions.   Indeed, as mentioned in Chapter 1, there are four dimensional asymptotically locally anti-de Sitter vacuum black holes  with horizon topology that of a surface of arbitrary genus.  Higher dimensional versions
of these ``topological" black holes have been considered, for example, in \cite{Bir, Mann}.  

Nevertheless, as Gibbons pointed out in \cite{Gi}, although Hawking's theorem does not hold in the asymptotically locally anti-de Sitter setting,  his basic argument still
leads to an interesting conclusion.  Gibbons showed that for a time-symmetric ($K=0$) spacelike hypersurface $V$ in a four dimensional spacetime $M$ satisfying the Einstein equation 
\eqref{eeq+}, such that $\calT$ obeys the DEC and $\Lambda < 0$, an
outermost MOTS  $\S$ (which in this case is a minimal surface)  must satisfy the area bound,
\beq\label{areabound}
{\rm Area}(\S)\ge \frac{4\pi(g-1)}{|\Lambda|} \, ,
\eeq
where $g$ is the genus of $\S$.   Woolgar \cite{Wo}  obtained a similar bound in  the general, nontime-symmetric, case.
Hence, for stationary black holes in this setting, black hole entropy has a lower bound depending on a global topological invariant, namely, the Euler characteristic, $\chi_{\S} = 2-2g$.

In \cite{CG},  Gibbon's result was extended to higher dimensional spacetimes.  There  it was shown, in the time-symmetric case, that a bound similar to that obtained by Gibbons still holds, but where the genus is replaced by the so-called $\s$-constant (or Yamabe invariant).  The $\s$-constant is an invariant of smooth compact manifolds that in dimension two reduces to a multiple of the Euler characteristic.  More recently, it was shown in \cite{GO} that, by using arguments similar to those used here to generalize Hawking's black hole topology theorem, this lower area bound can be extended to the nontime-symmetric case. We take a moment to describe this result.   

We begin by recalling the definition of the $\s$-constant.  
Let $\S^{n-1}$, $n\ge 3$, be a smooth compact (without boundary) $(n-1)$-dimensional
manifold.  If $g$ is a Riemannian metric on $\S^{n-1}$, let $[g]$ denote the class of metrics conformally related to $g$.  The Yamabe constant with respect to $[g]$, which we denote by 
$\calY[g]$, is the number, 
\beq\label{yam}
\calY[g]  = \inf_{\tilde g\in [g]} 
\frac{\int_{\S}S_{\tilde g}d\mu_{\tilde g}}
{(\int_{\S}d\mu_{\tilde g})^{\frac{n-3}{n-1}}}\, ,
\eeq  
where $S_{\tilde g}$ and $d\mu_{\tilde g}$ are respectively the scalar curvature and volume measure of $\S^{n-1}$ 
in the metric $\tilde g$.  The  expression involving integrals is just the volume-normalized total
scalar curvature of $(\S,\tilde g)$.
The solution to the Yamabe problem, due to Yamabe, Trudinger, Aubin and Schoen, 
guarantees that the infimum 
in (\ref{yam}) is  achieved by a metric of constant scalar curvature.  

The $\s$-constant  of $\S$ is 
 defined by taking the supremum of the Yamabe constants over all conformal
classes,
\beq
\s(\S) = \sup_{[g]} \calY[g] \, .
\eeq    
As observed by Aubin, the supremum is finite, and in fact bounded above in terms of the volume
of the standard unit $(n-1)$-sphere $S^{n-1} \subset \Bbb R^n$.  The $\s$-constant divides
the family of compact manifolds into three classes according to: (1) $\s(\S) > 0$, (2) $\s(\S) = 0$,
and (3) $\s(\S) < 0$.  

In the case $\dim \S =2$, the Gauss-Bonnet theorem implies $\s(\S) = 4\pi\chi(\S)=8\pi(1-g)$.  
Note that the inequality \eqref{areabound} only gives information when $\chi(\S) < 0$.
Correspondingly, in higher dimensions, we shall only be interested in the case 
when $\s(\S) < 0$.  It follows from the resolution of the Yamabe problem that
$\s(\S) \le 0$ if and only if $\S$ does not carry a metric of positive scalar curvature.
In this case, and with $\dim \S = 3$, Anderson \cite{An} has shown, as an application of Perlman's work on the geometrization conjecture, that  if $\S$ is hyperbolic, i.e.  carries a metric  of
constant curvature $-1$, then the $\s$-constant is achieved for this metric and $\s(\S) < 0$.


Turning to the spacetime setting, we have the following result.  

\begin{thm}[\cite{GO}]\label{volbound}  
Let $V^n$, $n \ge 4$, be a spacelike hypersurface in a spacetime 
satisfying the Einstein equations \eqref{eeq+}, such that the fields giving rise to $\calT$ obey the DEC and $\Lambda < 0$.   Let $\S^{n-1}$ be an outermost MOTS  in $V^n$ such that $\s(\S) < 0$.
Then the $(n-1)$-volume of $\S$ satisfies,
\beq
{\rm vol}(\S^{n-1}) \ge \left(\frac{|\s(\S)|}{2|\Lambda|}\right)^{\frac{n-1}2} \, .
\eeq   
\end{thm}

In  fact for this result, it is sufficient that $\S$ be stable.  We refer the reader to \cite{GO} for
further details.

\section{Further constraints on black hole topology}
\label{further}

In Section \ref{obstruct} we briefly describe some of the major developments  in the study of manifolds of 
positive scalar curvature which have led to restrictions on the topology of  manifolds of positive Yamabe type.  In Section \ref{cobordant} we consider some restrictions on horizon topology in six dimensional black holes arising from cobordism theory.  For results on $4$-manifolds referred to below, see, e.g., \cite{Scorpan}.

\subsection{Remarks on obstructions to the existence of  positive scalar \\curvature metrics}
\label{obstruct}

As we have shown, for spacetimes obeying the dominant energy condition, outermost
MOTSs, in particular, cross sections of the event horizon in stationary black holes, must admit  metrics of positive scalar curvature.   In Section \ref{gen} we described how this requirement restricts
the horizon topology of five dimensional black holes. 
The general problem of determining obstructions to the existence of positive scalar curvature metrics is one that has been studied for many years.  The first major result in this direction in higher dimensions is due to  Lichnerowicz \cite{Lich}.   Using a Bochner-type argument, he showed that if $\S$ is a compact spin manifold with a metric of positive scalar curvature then the kernel and cokernel of the Dirac operator vanish. In particular if $\S$ has dimension $4k$, then the so-called $\hat A$-genus,
which agrees with the index of the Dirac operator, must vanish, $\hat A(\S) = 0$.  In the case $\S$ is four dimensional, the $\hat A$-genus is related to the intersection form  
$Q_{\S} : H^2(\S;\bbZ) \times H^2(\S;\bbZ) \to \bbZ$,\,\footnote{By de Rham's theorem, modulo torsion, classes $\a, \b \in H^2(\S;\bbZ)$ can be represented by $2$-forms $\a^*, \b^*$, and then
$Q_{\S}(\a,\b) = \int_{\S} \a^* \wedge \b^*$.}
 by, $\hat A(\S) = -\frac18  {\s}(\S)$, where 
${\rm \s}(\S)$ is the signature of $Q_{\S}$.   But, there are known to be infinitely many smooth compact spin four manifolds with nonzero signature, the $K3$ surface being one such example.   
Moreover, there are higher dimensional analogues of  the $K3$ surface which have nonzero $\hat A$-genus; see e.g., \cite[p. 298]{Lawson}.  As a consequence, all of these examples fail to admit metrics of positive scalar curvature.  In \cite{Hitch} Hitchin generalized the vanishing theorem of  Lichnerowicz and obtained the surprising result that in every dimension $k>8$, there are smooth manifolds $\S^k$ homeomorphic to the standard sphere $S^k$, that do not admit metrics of positive scalar curvature; these manifolds must be exotic spheres. 

While these results are quite striking, they left open some very basic questions, for example, the question as to whether the $k$-torus, $k \ge 3$, admits a metric of positive scalar curvature. 
Then in \cite{SYsc}, Schoen and Yau made a major advance by proving, using minimal surface techniques, that if the fundamental group of a compact orientable $3$-manifold  contains a subgroup isomorphic to the fundamental group of a  surface of genus $g \ge 1$
then the manifold does not admit a metric of positive scalar curvature.  Hence, in particular,
the $3$-torus does not admit a metric of positive scalar curvature.   In \cite{SYsc2}, Schoen and Yau generalized their techniques to higher dimensions, thereby establishing inductively the existence of a large class of compact manifolds, including tori, of dimension up to~$7$, that do not admit metrics of positive scalar curvature.   The fundamental observation made in   \cite{SYsc2} is that if $\S^n$,  $3 \le n \le 7$, is a compact orientable manifold of positive scalar curvature then any nontrivial 
codimension-one homology class can be represented by a manifold that admits a metric of positive scalar curvature.  This is proved by choosing a manifold of least area in the homology class, and making use of the positivity of the minimal surface stability operator, ``rearranged"  in an especially useful way.

Another very important development in the study of manifolds of positive scalar curvature was
Gromov and Lawson's introduction  of the notion of {\it enlargability} \cite{GL1, GL2, GL3}.   In \cite{GL3} they extended their methods to noncompact manifolds, which required an adaptation of Dirac operator methods to noncompact manifolds, and which enabled them to strengthen some of their previous results.  For example, using these improved techniques they 
were able to show that a compact $3$-manifold which has a $K(\pi, 1)$ factor in its prime decomposition cannot admit a metric of positive scalar curvature, a result we used in Section \ref{gen}.
They also proved the following.
 
\begin{thm}[\cite{GL3}]  A compact manifold of arbitrary dimension which admits a metric of nonpositive sectional curvature, cannot admit a metric of positive scalar curvature.
\end{thm}

In the context of the results discussed in Sections \ref{gen} and \ref{borderline}, this rules out many obvious horizon topologies, including tori of all dimensions.

Finally, we mention that Seiberg-Witten theory provides further examples of
 compact simply connected $4$-manifolds that do not admit metrics of positive scalar curvature.
This relies on the following vanishing theorem (see, e.g. \cite{Scorpan}):  If $\S$ is a compact
$4$-manifold with $b_2^+(\S) \ge 2$, where $b_2^+(\S)$ is the number of positive eigenvalues
of the intersection form of $\S$ (or, equivalently, the dimension of the space of self-dual harmonic $2$-forms), and $\S$ admits a metric of positive scalar curvature then
the Seiberg-Witten invariants of $\S$ vanish.   The proof is again a Bochner type argument,
now based on the {\it coupled} Lichnerowicz equation, which, in addition to the scalar curvature
of $\S$, includes a term involving the curvature of the connection on the determinant line bundle of the specified complex spin structure.  At the same time, there are well-known classes of compact
simply connected $4$-manifolds that have nonvanishing Seiberg-Witten invariants 
(see e.g.,~\cite{Scorpan}).

\subsection{Cobordism constraints on four dimensional horizons}
\label{cobordant}

Consider a six dimensional asymptotically flat black hole spacetime $M$.  One can imagine (and even construct under suitable circumstances) a smooth spacelike hypersurface $V$ in $M$ that meets the
event horizon $H$ in a smooth compact $4$-manifold $\S$ and extends out to spatial infinity.  
In this situation $\S$ will be cobordant to a large sphere $\S' \approx S^4$ out near infinity.  That is, there is a compact region $W$ in $V$ whose boundary $\d W$ is the (appropriately oriented) union of
$\S$ and $\S'$.  In \cite{Oz}, the authors examine, among other things, the consequences of $\S$ being cobordant to
a $4$-sphere, while taking advantage of the classification theorem of Freedman and subsequent
work of Donaldson.  (See \cite{Scorpan} for a nice exposition of these  results.) 
For this discussion, which refines that in \cite{Oz}, it is assumed that $\S$ is simply connected.

In addition to these classification results, the key fact is the following:  If two compact oriented $4$-manifolds are cobordant then they have the same signature.  Hence, since $\S$ is cobordant
to $S^4$, $\s(\S) = 0$.

Now, from Freedman's classification  of simply connected $4$-manifolds in terms of intersection forms, the algebraic classification of intersection forms (as symmetric bilinear unimodular forms), and restrictions on smooth $4$-manifolds imposed by Donaldson's work we know the following
\cite{Scorpan}:

\begin{thm} Every smooth compact simply connected $4$-manifold is homeomorphic to $S^4$
or to one of the following connected sums,

\medskip
\noindent
(i)  $(\#m \bbC\bbP^2) \, \# \, (\#n\overline{\bbC\bbP^2})$.

\medskip
\noindent
(ii)
$(\#m \, S^2 \times S^2) \, \#\, (\#n \calE_8)$ or $(\#m \,S^2 \times S^2) \, \#\, (\#n(- \calE_8))$.
\end{thm}

Recall, $\bbC\bbP^2$ is the complex projective plane, $\overline{\bbC\bbP^2}$ is the same, but with the opposite orientation, $\calE_8$ is the $4$-manifold discovered by Freedman with intersection form the $E_8$ lattice, and $-\calE_8$ is the same, but with the opposite orientation. 

Using the fact that the signature is additive with repect to connected sum,
$$
\s(\S_1 \# \S_2) = \s(\S_1) + \s(\S_2) \,,
$$
along with the basic signature values, $\s(\bbC\bbP^2) = 1$, $\s(\overline{\bbC\bbP^2}) = -1$, $\s(\pm \calE_8) =  \pm 8$, and $\s(S^2 \# S^2) =  0$, we see that the vanishing of the signature of $\S$ implies  in (i) above that $m = n$, and in (ii) above that $n = 0$.  

We conclude that for our cross-section of the event horizon $\S$, if it is simply connected as well as cobordant to $S^4$, it  must be homeomorphic to $S^4$, or to a finite connected sum of 
 $S^2 \times S^2$'s or to a finite connected sum of $\bbC\bbP^2 \# \overline{\bbC\bbP^2}$'s.
In particular, $\S$ cannot be homeomorphic to $\bbC\bbP^2$ (as observed in \cite{Reall}) or to a $K3$ surface. 

We obtain further restrictions on the topology of $\S$ if we assume in addition that $\S$ is a spin manifold. 
This would be the case, for example if the spacelike hypersurface $V$ is spin, for then $\S$ 
would inherit a spin structure from $V$.   If $\S$ is spin then its
second Stiefel-Whitney class $w_2 \in H^2(\S,\bbZ_2)$, the obstruction to being spin, must  vanish.  In turn it follows from Wu's formula~\cite{Scorpan} that the intersection form of $\S$ is even, i.e., for all classes $\a$, $Q_{\S}(\a,\a)$ is even.  This rules out the connected sums of $\bbC\bbP^2 \# \overline{\bbC\bbP^2}$'s, and we finally arrive at: {\it The cross-section of the event horizon $\S$, if it is simply connected and spin, as well as cobordant to $S^4$,  must be homeomorphic to $S^4$ or to a finite connected sum of  $S^2 \times S^2$'s.}

It is worth noting that black hole dimension six is the first dimension where cobordism theory 
becomes useful.
This is due to the nontrivial fact, relevant to five dimensional black holes, that any two compact $3$-manifolds are (oriented) cobordant.  In addition, it is a fact that the region ``filling in" 
two $k$-dimensional (oriented) cobordant manifolds, $k \ge 3$,  can be taken to be simply connected. It is for this reason that topological censorship, which would imply in the discussion above that $\S$ is simply connected, does not provide any general constraints on horizon topology in five or higher spacetime dimensions.\footnote{An exception to this can occur when symmetries permit dimensional reduction; cf., \cite{CGS, Hollands2}.}

\section{Final remarks}
\label{final}

In Section \ref{obstruct} we focused on negative results concerning the existence of metrics of positive
scalar curvature.  There are also many positive results.  For instance, there are the ``gluing" (or surgery) results obtained independently, and by different methods, by Schoen-Yau \cite{SYsc2} and Gromov-Lawson \cite{GL2}, which show, in particular, that the connected sum of manifolds admitting metrics of positive scalar curvature admits a metric of positive scalar curvature. That is, positive Yamabe type is preserved under connected sums.   Using these surgery results and techniques and results from cobordism theory, Gromov and Lawson \cite{GL2} were able to prove the following.  Suppose $\S$ is compact, simply connected, with dimension $k\ge 5$. 

\medskip
(a)  If $\S$ is not spin then $\S$ admits a metric of positive scalar curvature.

\smallskip
(b)  If $\S$ is spin and is spin cobordant to a manifold which admits a metric 
\\\hspace*{.35in} of positive scalar  curvature that $\S$ admits a metric of positive scalar
\\\hspace*{.35in} curvature.

\smallskip
Since the spin cobordism groups are trivial in dimensions
$k = 5, 6 ,7$, it follows that every compact simply connected manifold of dimension 5, 6 or~7
admits a metric of positive scalar curvature.   

We make one further completely elementary observation:  Any manifold of the form 
$S^k \times M$, where $k \ge 2$ and $M$ is any compact manifold, admits a metric of positive scalar curvature.  Indeed, if $g'$ is a round metric on $S^k$ of radius $r$ and $g''$ is any metric on $M$ then the product metric $g' \oplus g''$ will have positive scalar curvature provided one takes
$r$ sufficiently small.  

Thus, while the requirement that the horizon be of positive Yamabe type puts rather strong  restrictions on the  topology of three dimensional horizons (as discussed in Section \ref{gen}), the situation becomes considerably more flexible in higher dimensions.   Indeed, the constraints on horizon topology described in this article still allow for a wide variety of possible topologies.  The 
method of blackfolds discussed in Chapter 8 provides an approach to realizing
many such topologies; see also \cite{fernando}.  The higher dimensional near-horizon geometries constructed, for example, in \cite{Lucietti}, many of which satisfy both constraints of being positive Yamabe and cobordant to spheres,  suggest even more possibilities.

\providecommand{\bysame}{\leavevmode\hbox to3em{\hrulefill}\thinspace}
\providecommand{\MR}{\relax\ifhmode\unskip\space\fi MR }
\providecommand{\MRhref}[2]{%
  \href{http://www.ams.org/mathscinet-getitem?mr=#1}{#2}
}
\providecommand{\href}[2]{#2}

\end{document}